\documentclass{article}
\usepackage{spconf,amsmath,graphicx}
\usepackage[fontsize=9pt]{scrextend}
\usepackage{amsmath,epsfig,amssymb,amsthm,url}
\usepackage{algorithm}
\usepackage{algorithm,algpseudocode}
\usepackage{multirow}
\usepackage{mdframed}
\usepackage{url}
\usepackage{array}
\usepackage{epstopdf}
\usepackage{float}
\usepackage{subfig}
\usepackage{breqn}
\usepackage{cite}
\usepackage{balance}

\newcommand{\argmin}{\operatornamewithlimits{argmin}}

\title{New Interpretable Patterns and Discriminative Features from \\Brain Functional Network Connectivity Using Dictionary Learning
}
\name{F. Ghayem$^*$, H. Yang$^*$, F. Kantar$^*$, S.-J. Kim$^*$, V. D. Calhoun$^{**}$, T. Adali$^*$ \thanks{This work was supported in part by NSF-NCS 1631838, and NIH grants R01 MH118695, R01 MH123610, R01 AG073949. The hardware used in the computational studies is part of the UMBC High Performance Computing Facility (HPCF).}}

\address{$^*$ Dept. of CSEE, University of Maryland Baltimore County, Baltimore, USA\\
	$^{**}$ Tri-institutional Center for Translational Research in Neuroimaging and Data Science (TReNDS), \\
	Georgia State University, Georgia Institute of Technology, and Emory University, Atlanta, USA
	}    
    
\newcommand{\Rbb}{{\mathbb{R}}}

\newcommand{\Ab}{{\bf A}}

\newcommand{\lb}{{\bf l}}
\newcommand{\Lb}{{\bf L}}

\newcommand{\db}{{\bf d}}
\newcommand{\fb}{{\bf f}}
\newcommand{\Fb}{{\bf F}}

\newcommand{\sbb}{{\bf s}}

\newcommand{\xb}{{\bf x}}

\newcommand{\zb}{{\bf z}}
\newcommand{\Zb}{{\bf Z}}

\newcommand{\Db}{{\bf D}}

\newcommand{\Wb}{{\bf W}}

\newcommand{\nf}[1]{\|#1\|_F} 
\newcommand{\nt}[1]{\|#1\|_2}

\newcommand{\ub}{{\mathbf u}}

\newsavebox\mybox

\makeatletter
\newcommand{\Diag}[1]{\mathop{\operator@font Diag}\{#1\}}
\renewcommand{\vec}{\mathop{\operator@font vec}}   %
\newcommand{\vecT}{\mathop{\operator@font vecT}}   %
\newcommand{\Unvec}{\mathop{\operator@font Unvec}}   %
\newcommand{\Span}{\mathop{\operator@font span}}
\makeatother

\makeatother

\newtheorem{mydef}{Definition}

\usepackage{color}
\definecolor{darkred}{rgb}{0.7,0,0}
\definecolor{darkbrown}{RGB}{110,75,45}
\definecolor{darkgreen}{rgb}{0,0.46,0}

\begin{document}
%
\maketitle
\begin{abstract}
%
%
Independent component analysis (ICA) of multi-subject functional magnetic resonance imaging (fMRI) data has proven useful in providing a fully multivariate summary that can be used for multiple purposes. ICA can identify patterns that can discriminate between healthy controls (HC) and patients with various mental disorders such as schizophrenia (Sz). Temporal functional network connectivity (tFNC) obtained from ICA can effectively explain the interactions between brain networks. On the other hand, dictionary learning (DL) enables the discovery of hidden information in data using learnable basis signals through the use of sparsity.
In this paper, we present a new method that leverages ICA and DL for the identification of directly interpretable patterns to discriminate between the HC and Sz groups. We use multi-subject resting-state fMRI data from $358$ subjects and form subject-specific tFNC feature vectors from ICA results. Then, we learn sparse representations of the tFNCs and introduce a new set of sparse features as well as new interpretable patterns from the learned atoms. Our experimental results show that the new representation not only leads to effective classification between HC and Sz groups using sparse features, but can also identify new interpretable patterns from the learned atoms that can help understand the complexities of mental diseases such as schizophrenia.

\end{abstract}


\section{Introduction}\label{sec:intro}
An important goal in neuroscience is the development of methods for identifying interpretable patterns that provide discrimination between healthy controls (HC) and different groups of patients. Functional magnetic resonance imaging (fMRI) has proven useful for the study of healthy brain as well as different brain disorders, including schizophrenia (Sz) \cite{fmri1, fmri2, fmri3}.
However, the high dimensionality structure and the noisy nature of fMRI data cause some challenges. Independent component analysis (ICA) has proven particularly useful for feature selection and statistical analysis of fMRI data \cite{SPM2014,ica1, ica3}.

ICA is a data-driven approach that decomposes the fMRI data into a set of independent components and their corresponding time courses (TC). Unlike classical model-driven methods, such as the general linear model (GLM)\cite{GLM} that requires predefined model parameters, ICA decomposes brain activities into functional networks that are maximally independent. By concatenating individual subject data and applying one ICA estimation on the aggregate data, group ICA (GICA) \cite{calhoun2001method} generalizes ICA to multi-subject analysis that provides for group inferences. The brain networks that are identified by ICA can be used for studying the inter-network relationships through temporal functional network connectivity (tFNC): the Pearson correlations between pairs of TCs. TFNC has been shown to be highly informative, for instance, chronic mental disorders such as schizophrenia are characterized by significant abnormalities in brain connections \cite{feat-fnc1,FNC2}.


%
Deep learning methods are also frequently used to detect discriminating biomarkers, but the findings are not directly interpretable and additional steps such as relevance propagation are typically needed \cite{montavon2018methods, NN1, NN2, NN3}.
In contrast, matrix decompositions such as dictionary learning offer a sparse linear decomposition of the signals based on a set of interpretable bases called atoms \cite{DL-Elad}.
These techniques have shown to extract new, easily comprehensible patterns from the data, revealing data’s hidden information \cite{ComoJ10, LustDSP08, CandW08, EldaG12, FoucR13}.

In this paper, we develop a method to extract a set of powerful features from resting state fMRI data by bringing together the advantages of ICA and DL. Given that tFNCs are effective in differentiating between the HC and Sz groups, these new features are derived from the sparse representation of the tFNCs from resting-state fMRI data. Our experimental findings show that compared with the original tFNCs, these new features help improve the classification performance between HC and Sz groups. In addition, the approach identifies novel, interpretable biomarkers that help explain the complexities of brain disorders such as schizophrenia.

In the rest of this paper, Section~\ref{sec:review} reviews the ICA and DL methods. The proposed framework is discussed in Section~\ref{sec:proposed}.
Section~\ref{sec:simulations} presents experimental results. The conclusions and perspectives are discussed in Section~\ref{sec:Conclusion}.

\section{Background}\label{sec:review}
In this section, we first provide a background on how the functional network connectivity features we make use of are created. Then, we review dictionary learning, a central element of our method.
\subsection{Functional network connectivity using group ICA}\label{subsec:EBM}
Consider the mixture model $\xb(v) = \Ab \sbb(v)$, where $\xb(v)$ is the observed mixture of $N$ statistically independent signals (components) $\sbb(v) = \left[ s_1(v) \dots s_N(v)\right]^T$ at voxel $v$ mixed via $\Ab$.
The components can be estimated as $\hat{\sbb}(v) = \Wb \xb(v)$, where $\Wb\in \Rbb^{N\times N}$ is a demixing matrix, which can be obtained by ICA.
%
%
We make use of GICA \cite{G-ICA}, where temporal concatenation of subject datasets is applied to form group data followed by performing one ICA on the group data. The subject-specific result can be achieved by back-reconstruction, which allows the comparison of spatial maps and time courses across subjects, while also addressing the issue with permutation ambiguity inherent in the ICA \cite{calhoun2009review}.
The temporal interactions between brain networks for the $k^{\text{th}}$ subject can be presented as $\text{tFNC}^{[k]}\in \mathbb{R}^{N \times N}$ through the Pearson correlation between time courses from  $\Ab^{[k]}$, which underwent postprocessing procedure in \cite{allen2014tracking}. Since $\text{tFNC}^{[k]}$ is symmetric, for subject $k$, we only use its upper triangular data, and arrange them in vector $\fb^{[k]}$ of size $P = \frac{N(N-1)}{2}$ to serve as the subject's \textit{FNC-feature} vector.
\subsection{Dictionary learning and sparse representation}
\label{ssec:DL}
Dictionary learning is the task of estimating a set of basis signals, called atoms, using a training dataset such that each training sample can be written as a sparse linear combination of the learned atoms. Mathematically, 
consider some training feature vectors $\fb^{[k]} = [f^{[k]}_1, \dots, f^{[k]}_P]^T$ collected as the columns of the matrix $\Fb = [\fb^{[1]},\dots,\fb^{[K]}] \in \mathbb{R}^{P\times K}$. Then, the goal is to learn a dictionary $\Db = [\db_1, \dots, \db_G] \in \mathbb{R}^{P\times G}$, with G as the number of atoms, such that $\Fb = \Db \Zb$ and the columns of the coefficient matrix $\Zb = [\zb^{[1]},\dots,\zb^{[K]}]\in \mathbb{R}^{G\times K}$ are sparse.
That is, each feature vector $\fb^{[k]}$ can be written as $\fb^{[k]} = \sum^{G}_{i=1} z_g\db_g$, where most of $z_g$'s are zeros.
To learn $\Db$, a sparsity promoting function denoted $r$ is considered to impose a sparsity constraint on the coefficient matrix $\Zb$. The dictionry learning problem is then formulated as follows \cite{DL-Elad}:
\begin{align}
\min_{\Db,\Zb}~
&\frac{1}{2}\nf {\mathbf{F} 
- \Db \Zb}^2
+\lambda \cdot r(\Zb),
\nonumber
\\
\mbox{s.t.}~~&
    \Db \in \mathcal{D} := \{ \Db: \nt{\db_g} = 1 , g=1,2,\dots,G \},
    \nonumber
    \tag{P1}
    \label{eq:DL}
\end{align}
where $\nf{.}$ is the Frobenius norm, and $\lambda >0$ is a sparsity parameter.

Problem \eqref{eq:DL} can be solved using alternating minimization, which iterates between two steps: dictionary update (DU), and sparse representation (SR).
The DU step minimizes \eqref{eq:DL} across dictionary $\Db$ while assuming that $\Zb$ is fixed, and in (SR) step, \eqref{eq:DL} is solved with respect to $\Zb$ assuming $\Db$ is fixed.
Different approaches can be used to alternate between these two steps \cite{DL-Elad, IPP, ADL, l0soft}.

\section{Methodology}
\label{sec:proposed}
In this section, we propose the sparse representation of the FNC-features from fMRI data, and show how it reveals new interpretable patterns that can provide discrimination between HC and Sz. 

\begin{algorithm}[t!]
		\caption{\small Proposed method for solving \eqref{eq:DL-LC}}
		\label{alg:DL-LC-FNC}
		\begin{algorithmic}[1]
			\State \textbf{Inputs:} $\Fb_{\text{tr}}$, $\Fb_{\text{ts}}$, $\Lb_{\text{tr}}$, $ \kappa$, $G$, $\mu$, $ \beta $, $\text{Iter}_{\text{in}}$, $\text{Iter}_{\text{out}}$
			\State \textbf{Initialization:} $\Db$ $\&$ $\Wb$: DCT , $\Zb=[0]_{G\times K}$
			\For{$i = 1 : \text{Iter}_{\text{out}}$}
			\For{$j = 1 : \text{Iter}_{\text{in}}$}
			\State
			SR-gradient:
			\State
			~~~~
			$
                \nabla \mathbf{J}_{\text{Z}_{\text{ts}}} = 
                - \Db^T \Fb_{\text{ts}}
                +
                \Db^T \Db \Zb_{\text{ts}}
            $
            \State
            ~~~~
			$
                \nabla \mathbf{J}_{\text{Z}_{\text{tr}}}
                = 
                - \Db^T 
                (\Fb_{\text{tr}} - \Db \Zb_{\text{tr}})
                - \beta \Wb^T
                (\Lb_{\text{tr}} - \Wb \Zb_{\text{tr}})
            $
			\State 
			SR-sparsification:
			\State
			~~~~
			$
                \Zb_{\text{ts}}
                \leftarrow
                \text{prox}_{\mu\lambda r(\cdot) }
                (\Zb_{\text{ts}}-\mu\nabla \mathbf{J}_{\text{Z}_{\text{ts}}})
            $
            \State 
            ~~~~
			$
                \Zb_{\text{tr}}
                \leftarrow
                \text{prox}_{\mu\lambda r(\cdot) }
                (\Zb_{\text{tr}}-\mu\nabla \mathbf{J}_{\text{Z}_{\text{tr}}})
            $
			\EndFor
			\State 
			$
                \Zb
                \leftarrow
                [\Zb_{\text{tr}},\Zb_{\text{ts}}]
            $
			\State
			DU: 
			    $
                    \Db 
                    \leftarrow
                    \mathcal{P}_{\mathcal{D}}\Big(
                    \mathbf{F}
                    \Zb^T 
                    (\Zb \Zb^T )^{-1}
                    \Big)
                $
                \State
            Classifier update:
			$
                \Wb 
                \leftarrow
                \Lb_{\text{tr}}
                \Zb_{\text{tr}}^T 
                (\Zb_{\text{tr}} \Zb_{\text{tr}}^T )^{-1}
            $
			\EndFor
			\State \textbf{Output:} $\Db$, $\Zb$, $\Wb$
		\end{algorithmic}
	\end{algorithm}
\subsection{Joint classifier and dictionary learning for FNC-features}
We propose to jointly learn a linear classifier and a dictionary to find the sparse representations of the FNC-feature vectors $\fb^{[k]}=\Db\zb^{[k]}$.
We concatenate the feature vectors of all subjects in matrix $\Fb = [\fb^{[1]},\dots \fb^{[K]}]$, and consider a common dictionary $\Db \in \mathbb{R}^{P \times G}$ with $G$ as the number of atoms.
Then, the DL problem will be as in \eqref{eq:DL}.
To jointly learn a linear classifier with the dictionary $\Db$, we use the binary labels $\lb^{(\text{HC})}=[0,1]^T$ and $\lb^{(\text{Sz})}=[1,0]^T$ for HC and Sz groups, respectively.
So, with a linear classifier $\Wb\in\mathbb{R}^{2\times G}$, the label of subject $k$ can be estimated as $\lb^{[k]} = \Wb\zb^{[k]}$.
We denote the data corresponding to training and test sets with subscript ``tr'', and ``ts'', respectively.
By concatenating the labels for $K_{\text{tr}}$ training datasets in the label matrix $\Lb_{\text{tr}} = [\lb_{\text{tr}}^{[1]},\dots,\lb_{\text{tr}}^{[K_{\text{tr}}]}]$, we have $\Lb_{\text{tr}}=\Wb\Zb_{\text{tr}}$.
Inspired by \cite{zhang2010discriminative}, we add this equality as a discriminative penalty with parameter $\beta$ to the cost function in \eqref{eq:DL}, and solve the following problem:
\begin{align}
\min_{\Db,\Zb,\Wb}~
&\frac{1}{2}\nf {\mathbf{F} 
- \Db \Zb}^2
+\lambda \cdot r(\Zb)
+ \frac{\beta}{2}\nf {\Lb_{\text{tr}}
- \Wb \Zb_{\text{tr}}}^2,
\nonumber
\\
\mbox{s.t.}~~&
    \Db \in \mathcal{D} := \{ \Db: \nt{\db_g} = 1 , g=1,2,\dots,G \}.
    \nonumber
    \tag{P2}
    \label{eq:DL-LC}
\end{align}
We note that in the above formulation, we consider to update the dictionary $\Db$ using the whole dataset $\Fb=[\Fb_{\text{tr}},\Fb_{\text{ts}}]$ and $\Zb=[\Zb_{\text{tr}},\Zb_{\text{ts}}]$, while for the classification term, we only involve the training labels.

To solve \eqref{eq:DL-LC}, we perform alternating minimization.
At iteration $i$, dictionary update (DU) is obtained by setting the gradient of the target function with respect to $\Db$ to zero, and projecting the result to the set $\mathcal{D}$.
This results in a closed form expression
$
\text{(DU):}~~
    \Db^{(i)} 
    \leftarrow
    \mathcal{P}_{\mathcal{D}}\Big(
    \mathbf{F}
    \Zb^T 
    (\Zb \Zb^T )^{-1}
    \Big),
$
where $\mathcal{P}_{\mathcal{D}}$ is the projection on the set $\mathcal{D}$.

In the sparse representation (SR) step, different approaches such as orthogonal matching pursuit (OMP) and proximal methods can be used \cite{DL-Elad, IPP}.
Here, we use iterative proximal-projection approach due to its flexibility to a range of sparsity-promoting functions, including non-convex and non-smooth scenarios \cite{IPP}.
The \textit{proximal mapping} is a key operator in these algorithms defined as:
\begin{mydef}
\emph{\cite{PariB14}}
The proximal mapping of a proper and lower semicontinuous function $r: \emph{\text{dom}}_r \rightarrow (-\infty,+\infty]$ at $\xb\in\mathbb{R}^{n}$ is
$
\emph{\text{prox}}_r(\xb)
=
\argmin_{\ub\in \emph{\text{dom}}_r}~
\{
\frac{1}{2}\nf {\xb 
- \ub}^2
+ r(\ub)
\}.
\nonumber
$
\label{def:proximal}
\end{mydef}
We first decompose the sparse coefficients into training and testing sets, and we separately update $\Zb_{\text{tr}}$ and $\Zb_{\text{ts}}$ using proximal method.
The proximal approach consists of two steps: 1) gradient step, 2) sparsification step.
We start with updating $\Zb_{\text{ts}}$.
By defining $\mathbf{J}_{\text{Z}_{\text{ts}}} \triangleq \frac{1}{2}\nf {\Fb_{\text{ts}} - \Db \Zb_{\text{ts}}}^2$, the gradient step is
$
    \nabla \mathbf{J}_{\text{Z}_{\text{ts}}} = 
    - \Db^T \Fb_{\text{ts}}
    +
    \Db^T \Db \Zb_{\text{ts}}.
$
Then, the sparsification step is
$
\text{(SR-sparsification):}~~
    \Zb_{\text{ts}}
    \leftarrow
    \text{prox}_{\mu\lambda r(\cdot) }(\Zb_{\text{ts}}-\mu\nabla \mathbf{J}_{\text{Z}_{\text{ts}}}),
$
where $\text{prox}_{g}(\cdot)$ is given in Definition~\ref{def:proximal}.
Selecting various functions for $r(\cdot)$ results in different sparsifications.
For example, the SR-sparsification step will be soft-thresholding if we use the $\ell_1$-norm, and hard-thresholding if we use the $\ell_0$-norm.
In this paper, we consider a $\kappa$-sparse constraint, and set
\begin{equation}
r(\Zb_{\text{ts}}) \triangleq
\begin{cases}
 0 ~~~~&\text{if}~~ \|\Zb_{\text{ts}}\|_0 \leq \kappa \\
 \infty ~~~~ &\text{o.w.} 
\end{cases}.
\end{equation}
The proximal of the above function is a projection operation that keeps the $\kappa$ largest elements of $|\Zb_{\text{ts}}|$ with setting the rest to 0.

Now, we define
$\mathbf{J}_{\text{Z}_{\text{tr}}}
\triangleq
\frac{1}{2}\nf {\Fb_{\text{tr}} - \Db \Zb_{\text{tr}}}^2
+ \frac{\beta}{2}\nf {\Lb_{\text{tr}} - \Wb \Zb_{\text{tr}}}^2
$ for the update of $\Zb_{\text{tr}}$. With a similar procedure, we obtain the update as
\vspace{-9pt}
\setcounter{equation}{2}
\begin{equation}
\setlength{\jot}{-0pt} 
    \begin{aligned}
&\text{(SR-gradient):}
~~
    \nabla \mathbf{J}_{\text{Z}_{\text{tr}}}
    = 
    - \Db^T 
    (\Fb_{\text{tr}} - \Db \Zb_{\text{tr}})
    - \beta \Wb^T
    (\Lb_{\text{tr}} - \Wb \Zb_{\text{tr}}),
    \\
&\text{and (SR-sparsification):}
~~
    \Zb_{\text{tr}}
    \leftarrow
    \text{prox}_{\mu\lambda r(\cdot) }(\Zb_{\text{tr}}-\mu\nabla \mathbf{J}_{\text{Z}_{\text{tr}}}).
\end{aligned}
\nonumber
\end{equation}

\vspace{-9pt}
The update of the linear classifier $\Wb$ is obtained by setting the gradient of the target function in \eqref{eq:DL-LC} with respect to $\Wb$ to zero. This results in the closed form expression: $
    \Wb
    \leftarrow
    \Lb_{\text{tr}}
    \Zb_{\text{tr}}^T 
    (\Zb_{\text{tr}} \Zb_{\text{tr}}^T )^{-1}.
$
The final algorithm is summarized in Algorithm \ref{alg:DL-LC-FNC}.
\begin{figure}[t!]
  \centering
  \centerline{ \hspace{2mm}\includegraphics[scale=.33]{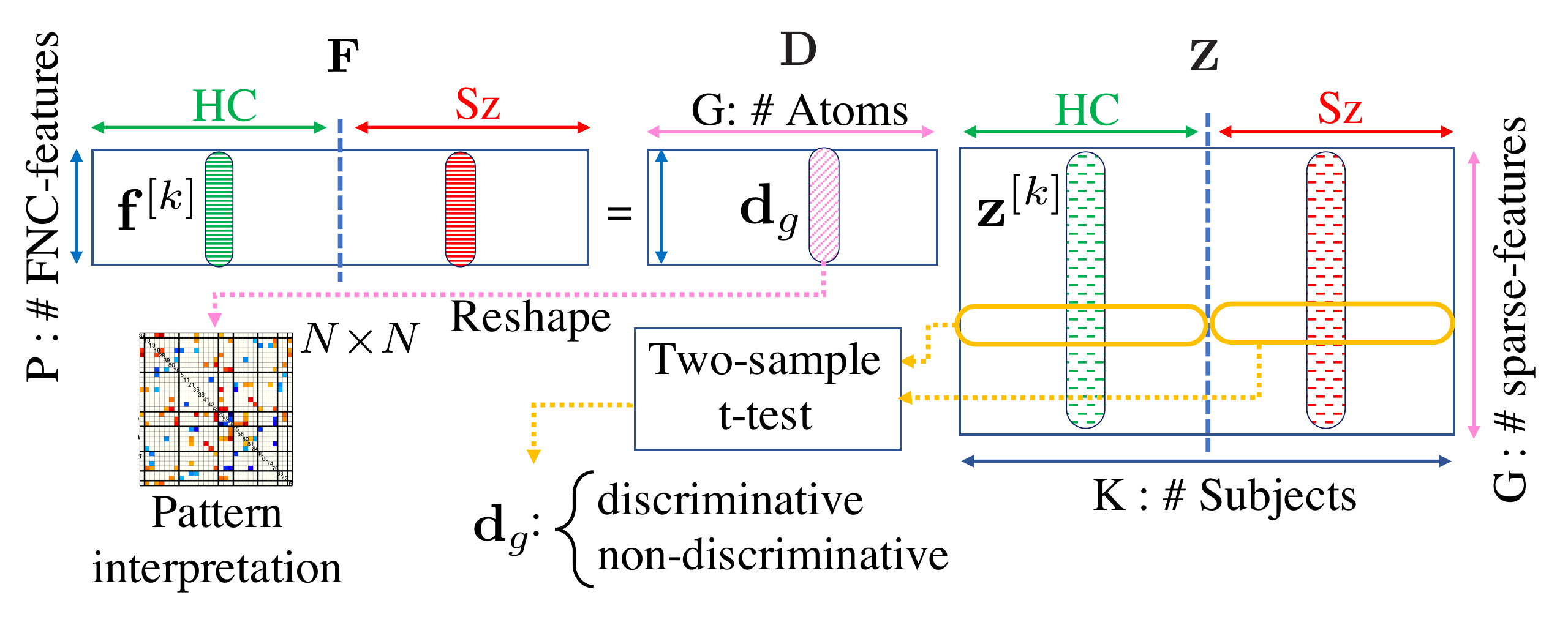}}
  \caption{Proposed framework.}
  \label{fig:interpret-discriminate}
	\end{figure}
\subsection{Interpretability, and discrimination in sparse space}
In this part, from the sparse decomposition of the tFNCs, we introduce new discriminative features, and new interpretable patterns.
To accomplish this goal, we suggest assigning the new \textit{sparse-feature} vector of subject $k$ as $\zb^{[k]}=[z^{[k]}_1,\dots,z^{[k]}_G]^T$, corresponding to the $k^{\text{th}}$ columns of $\Zb$.
Each element $z^{[k]}_g$ describes the contribution of the $g^{\text{th}}$ atom $\db_g$ to the representation of the initial tFNC-feature vector $\fb^{[k]}$.
On the other hand, every atom $\db_g$ has the same dimensionality as the original tFNC-feature vectors.
Therefore, by rearranging the atoms into a symmetric matrix of size $N\times N$, we can interpret the resulting matrices as in the tFNCs.
This interpretation again explains the interaction between brain networks $\{ \sbb^{[k]}_1,\dots,\sbb^{[k]}_N \}$ where $\sbb^{[k]}_n=[s^{[k]}_n(v)]^T$ for $v=1,\dots,V$ representing the voxels, but this time through the new patterns obtained from the atoms. 

Now, we can perform a statistical analysis to determine which of these atoms provide discrimination between the two groups. For example, if we divide the columns of the sparse coefficient matrix into the HC and Sz groups as $\Zb = [\Zb^{[\text{HC}]},\Zb^{[\text{Sz}]}]$, we can determine if the pattern corresponding to the $g^{\text{th}}$ atom is discriminating or not by performing a two-sample t-test \cite{student1908probable} between the $g^{\text{th}}$ rows of $\Zb^{[\text{HC}]}$ and $\Zb^{[\text{Sz}]}$.
Fig. \ref{fig:interpret-discriminate} illustrates the steps for interpretation as well as a comparison of the FNC-feature with the sparse-feature vector.
\section{Experimental results}
\label{sec:simulations}

\subsection{Data preparation}
\label{ssec:data-prep}
\noindent \textbf{\textit{Extraction of FNC-features.}}
We use multi-subject resting state fMRI (rs-fRMI) data from the bipolar and schizophrenia network for intermediate phenotypes (BSNIP) dataset \cite{tamminga2013clinical-et-al} considering 179 healthy controls (HC) and 179 patients with schizophrenia (Sz), using five sites: Baltimore, Chicago, Dallas, Detroit, and Hartford. All images were collected from a single 5-min run on a 3-T scanner and all subjects were instructed to have their eyes open and remain still during the entire scan. The fMRI data were then resampled to $3 \times 3 \times 3$ mm$^3$ isotropic voxels and smoothed using a Gaussian kernel with a full width at half maximum (FWHM) $=6$ mm. Only the subjects who passed quality control \cite{du2021evidence} were selected. We removed the first three timepoints for the following ICA analysis. Group ICA-EBM \cite{ICA-EBM} is performed to obtain the subject-specific tFNC-feature vectors $\fb^{[k]}$. The order is determined as $N=55$ using the method proposed in \cite{fu2014likelihood}. Compared with other ICA algorithms, ICA-EBM has the flexibility of estimating sources from different distributions by using a few classes of nonlinear functions.
Out of the 55 estimated components, we selected $N = 32$ as functionally relevant. The size of the tFNC-features was then calculated as $P=\frac{N(N-1)}{2}=496$.\footnote{The facility is supported by the U.S. National Science Foundation through the MRI program (grant nos. CNS-0821258, CNS-1228778, and OAC-1726023) and the SCREMS program (grant no. DMS-0821311).}

\noindent \textbf{\textit{Extraction of sparse-features.}}
We obtain subject-specific sparse-feature vectors $\zb^{[k]}$ by applying the DL approach presented in Section~\ref{ssec:DL} on the tFNC-features $\fb^{[k]}$. We initialize $\Db$ and $\Wb$ with DCT dictionary \cite{DL-Elad}, and $\Zb$ with a null (zero) matrix.
We consider a complete dictionary of size $G= P = 496$, and the sparsity level is set to $\kappa$-sparse $=50\%$.
The gradient descent step size is $\mu=0.005$, and the number of inner iteration and outer iteration are set to $\text{Iter}_{\text{in}}=5$, and $\text{Iter}_{\text{out}}=200$, respectively.
These values are selected empirically based on our observation regarding the convergence behaviour of the sparse representation of the signals.
We consider two different scenarios: 1) $\beta=0$ which represents DL without learning a linear classifier, and 2) $\beta=0.05$ which considers the linear classifier to be jointly learned with the dictionary.
We randomly select $20\%$ of the subjects within each group as test set $\Fb_{\text{ts}}$, and we keep the rest of the data ($80\%$) as the training set $\Fb_{\text{tr}}$.
With the above setup, we run Algorithm~\ref{alg:DL-LC-FNC} and obtain the sparse coefficients for the subjects in groups HC and Sz, \textit{i.e.} $\Zb^{[\text{HC}]}$ and $\Zb^{[\text{Sz}]}$.



\begin{table}[t!]
		\caption{Average classification rates $[\%]$.}
		\centering
		\resizebox{.45\textwidth}{!}{
		\begin{tabular}{|c|c|c|c|}
			\hline
			 Metric$\backslash$Feature &  tFNC & Sparse ($\beta=0$) & Sparse ($\beta=0.05$) \\
			\hline
			\hline
			Recall & $74.75\pm 0.61$ & $73.56\pm 0.65$ & $\mathbf{75.19}\pm 0.65$  \\
			Specificity & $73.78\pm 0.70$ & $74.14\pm 0.70$ & $\mathbf{74.47}\pm 0.68$  \\
			Precision & $74.35\pm 0.50$ & $74.27 \pm 0.53$ & $\mathbf{74.93} \pm 0.51$ \\
			Accuracy & $74.26\pm 0.40$ & $73.85 \pm 0.45$ & $\mathbf{74.83} \pm 0.43$  \\
			F1-score & $74.35\pm 0.40$ & $73.72 \pm 0.46$ & $\mathbf{74.87} \pm 0.45$  \\
			\hline
		\end{tabular}
		}
		\label{tab:classification}
	\end{table}
\subsection{Classification results}
In this section, we compare the performance of the tFNC-features and sparse-features in the classification of HC and Sz groups.
In order to achieve this, we train SVM classifiers \cite{svm} with polynomial kernels of order 3, which according to our experiments, provided the best overall performance.
In the training phase, we separately use features ($\Fb_{\text{tr}}$,$\Lb_{\text{tr}}$) and ($\Zb_{\text{tr}}$,$\Lb_{\text{tr}}$) to train SVM classifiers $\text{SVM}^{\text{FNC}}$ and $\text{SVM}^{\text{SPR}}$ using sparse-features and FNC-features, respectively.
Then, in the test phase, the test sets $\Fb_{\text{ts}}$ and $\Zb_{\text{ts}}$ are respectively given to $\text{SVM}^{\text{FNC}}$ and  $\text{SVM}^{\text{SPR}}$, in order to estimate the labels of the test sets $\widehat{\Lb}_{\text{ts}}^{\text{FNC}}$ and $\widehat{\Lb}_{\text{ts}}^{\text{SPR}}$.
Comparing the estimated test labels with the actual test label matrix $\Lb_{\text{ts}}$, we evaluate the classification performance using 5 metrics: recall, specificity, precision, accuracy, and F1-score.
By repeating the classification experiment 100 times with new random samples from the training and test sets, we report the average classification rates in Table~\ref{tab:classification}.
We see that all classification metrics are improved by using sparse features derived from the dictionary that are jointly learnt with the linear classifier.
In addition, as we address below, the decomposition provides better interpretability \textit{w.r.t.} Sz and HC differences.
\begin{figure}[t!]
  \centering
  \centerline{ \hspace{2mm}\includegraphics[scale=.2]{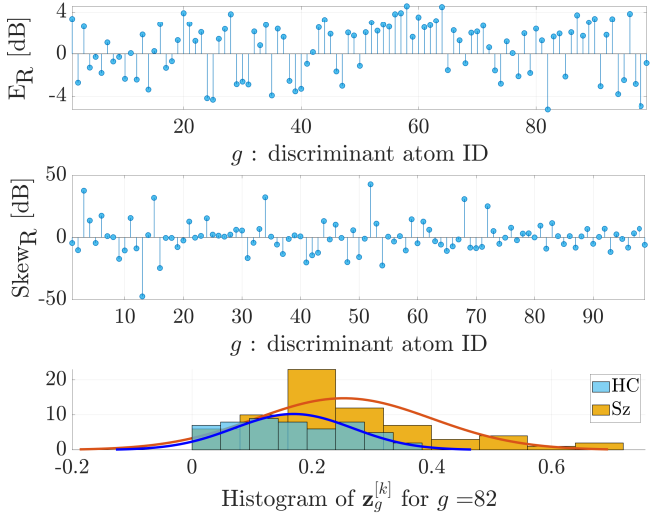}}
  \caption{Statistical analysis of the sparse coefficient corresponding to the 99 discriminative atoms obtained from the two-sample t-test between the sparse coefficients $\zb_g^{[\text{HC}]}$ and $\zb_g^{[\text{Sz}]}$. Top: the ratio between the average energy and, middle: the ratio between the skewness of $\zb_g^{[\text{HC}]}$ and $\zb_g^{[\text{Sz}]}$. Bottom: the histogram of $\zb_g^{[\text{HC}]}$ and $\zb_g^{[\text{Sz}]}$ for $g=82$.
  }
  \label{fig:stat-z}
	\end{figure}
\subsection{Discriminant atoms and their interpretability}
In order to find the discriminative atoms and tFNC-features between groups HC and Sz, we apply two-sample t-test followed by false discovery rate (FDR) correction \cite{benjamini2005false}.
For the FNC-features, the two-sample t-test is applied on the $p^{\text{th}}$ feature vectors $\fb^{[\text{HC}]}_p$ and $\fb^{[\text{Sz}]}_p$, which correspond to the subject indices in groups HC and Sz, respectively. We found 113 discriminative FNC-features out of $P=496$.
We repeated the two-sample t-test for the sparse coefficients corresponding to the $g^{\text{th}}$ atom $\zb^{[\text{HC}]}_g$ and $\zb^{[\text{Sz}]}_g$ (see Fig. \ref{fig:interpret-discriminate}), and we identified $99$ atoms (patterns) that discriminate between HC and Sz groups.

The sparse coefficients provide us with a statistical population that can be further analyzed for better understanding of the contribution of each discriminative atoms in the two groups.
In Fig. \ref{fig:stat-z}, the top plot shows the ratio between the average energy of the sparse coefficients for HC and Sz groups \textit{i.e.} $\text{E}_{\text{R}} = 10\log (||\zb^{[k_{\text{HC}}]}_g||_2^2 / ||\zb^{[k_{\text{Sz}}]}_g||_2^2)$.
Here, $g$ represents one of the $99$ discriminant atoms specified by the two-sample t-test.
The atoms with energy ratios above $0$ are those with higher energies for the HC group, which indicates that they dominant in the HC group, \textit{i.e.}, contribute more to the representation of tFNC-features for the HC group. The discriminative atoms that are dominant in Sz group are also indicated by points below the threshold of zero.
Also, a higher absolute value $|\text{E}_{\text{R}}|$ indicates a higher energy difference between the two groups.
The middle plot in Fig. \ref{fig:stat-z} reports the ratio between the skewness of the sparse coefficients for HC and Sz groups, \textit{i.e.}, $\text{Skew}_{\text{R}} = 10\log (\text{skew}(\zb^{[k_{\text{HC}}]}_g) / \text{skew}(\zb^{[k_{\text{Sz}}]}_g))$.
Similarly, atoms with positive values of $\text{Skew}_{\text{R}}$ are more skewed in the distribution of $\zb^{[\text{HC}]}_{g}$, and vice versa for the atoms with negative $\text{Skew}_{\text{R}}$.
A larger absolute value $|\text{Skew}_{\text{R}}|$ indicates more skewness difference between the two groups.
The bottom plot in Fig. \ref{fig:stat-z} shows the histogram of the sparse coefficients for a sample discriminative atom $g=64$. From the histogram, we can visually verify that the sparse coefficients corresponding to this atom have larger energy in Sz group and the skewness in the distribution of the sparse coefficients is higher in this group.

\begin{figure}[t!]
	\centering
	\hspace{-6mm}
	\subfloat[$\sum\db_g\mathbf{\bar{z}}^{[\text{Sz}]}_g$ for $g$'s dominant in Sz]{\includegraphics[scale=0.153]{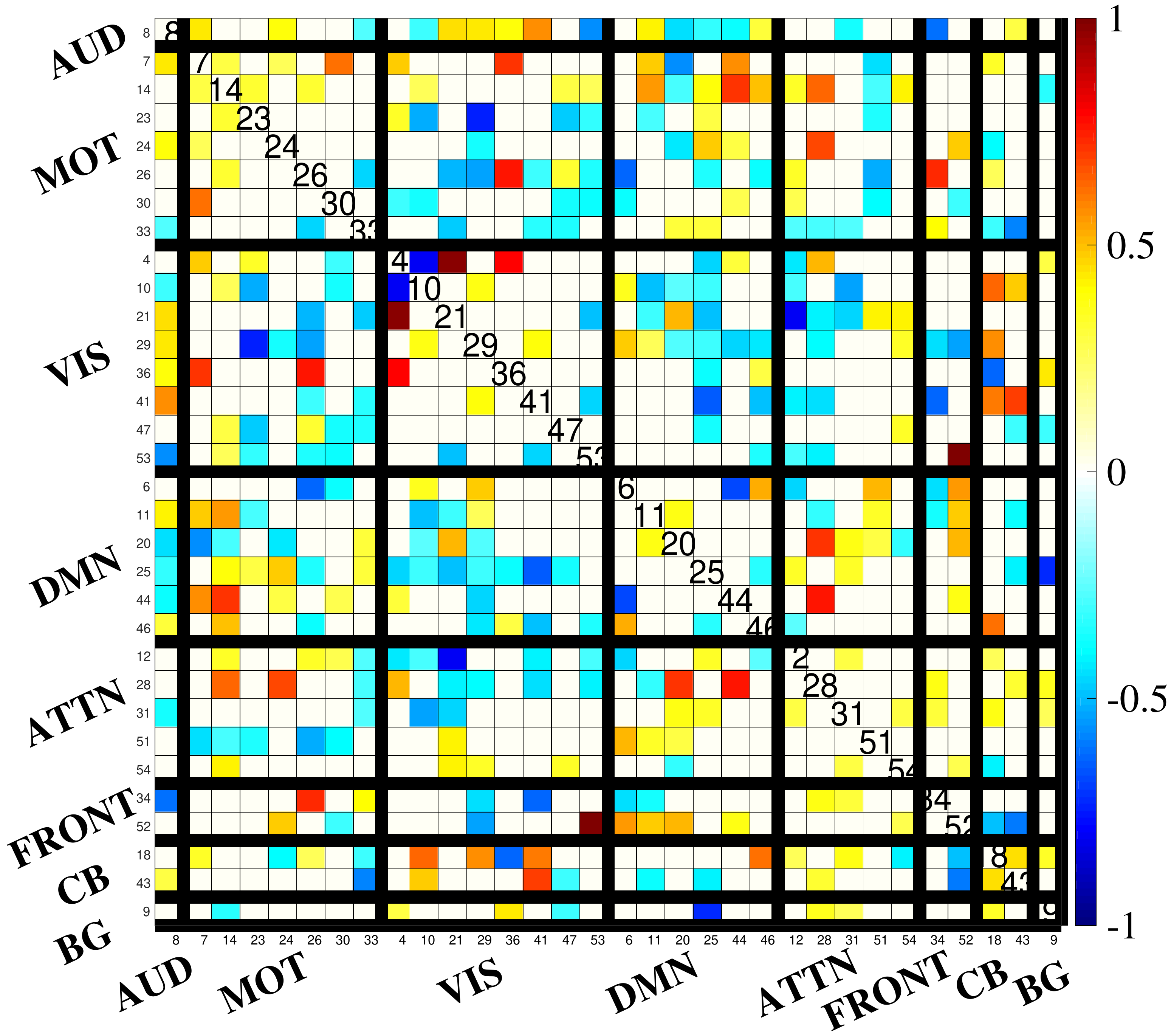}}
	\subfloat[$\sum\db_g\mathbf{\bar{z}}^{[\text{Sz}]}_g$ for $g$'s dominant in HC]{\includegraphics[scale=0.153]{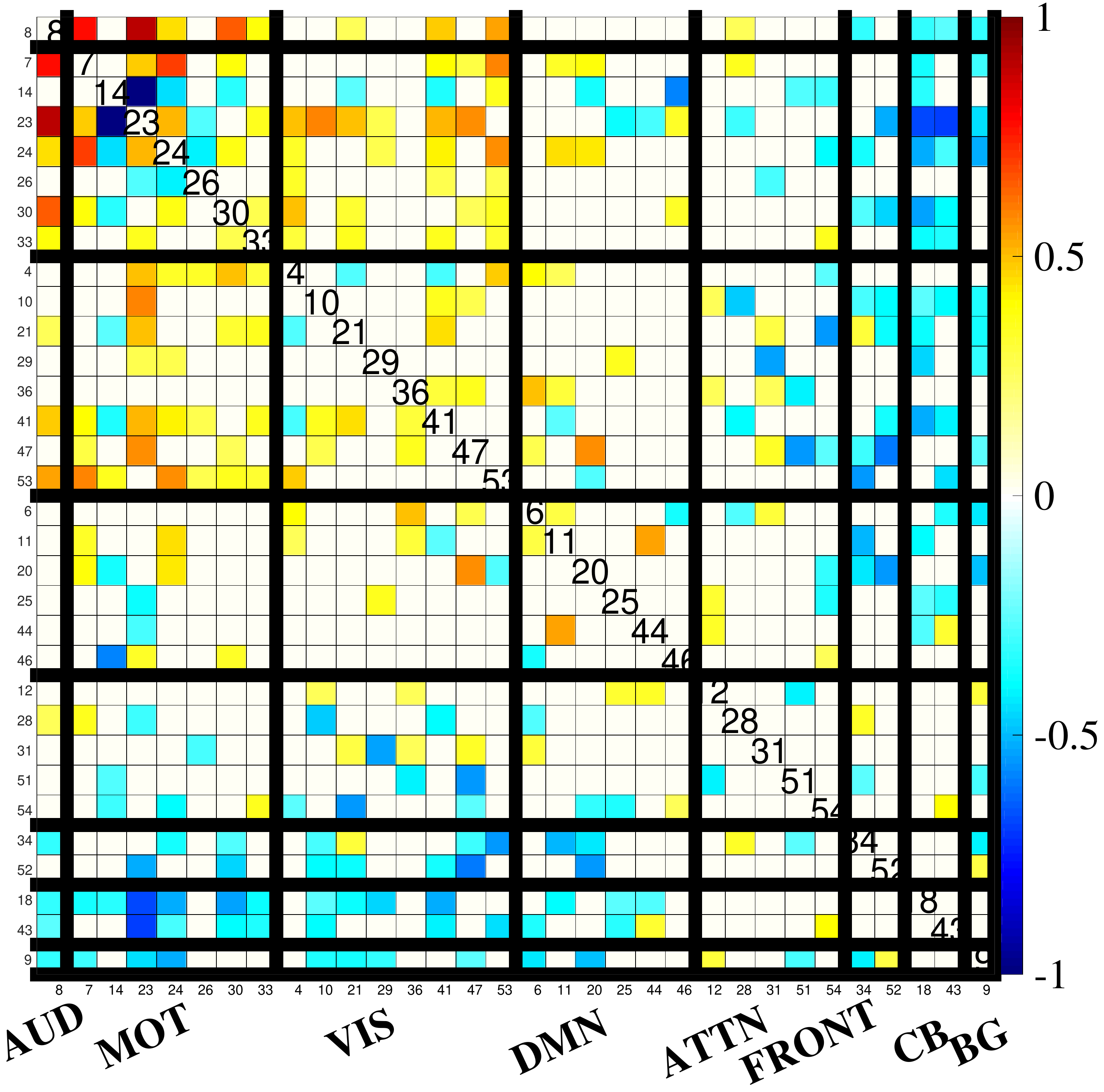}}
		\caption{Weighted average of the discriminative atoms that are a) dominant in HC, and b) dominant in Sz groups. More modularity is observed in HC.
		}
		\label{fig:sample-atom-reshaped}
\end{figure}

Besides information from the statistical analysis of the sparse coefficients, we can interpret the discriminative atoms by reshaping them to a symmetric matrix of size $N\times N$ which has the same shape as the tFNC.
Fig. \ref{fig:sample-atom-reshaped} shows the average pattern corresponding to the first 15 dominant atoms in Sz and HC groups with the largest energy ratios (according to Fig. \ref{fig:stat-z}-top). The results are shown with ae threshold level of $0.25$.
The pattern in Fig. \ref{fig:sample-atom-reshaped}-(a) constitutes $6.7\%$ of the overall energy in $\Zb^{[\text{HC}]}$, while it is only $3.5\%$ in $\Zb^{[\text{Sz}]}$. These contributions for Fig. \ref{fig:sample-atom-reshaped}-(b), are $1.3\%$ and $2.7\%$ for HC and Sz groups, respectively.
We note that there is a large difference in the energy between the two groups. This increases our confidence about the discrimination of these patterns between the two groups.
in Fig. \ref{fig:sample-atom-reshaped} we can see that these atoms show different patterns in terms of the interaction between different brain networks.
%
%
Comparing Fig. \ref{fig:sample-atom-reshaped} (a) and (b), we can see more modularity in HC with a swath of negative values in between sensory and DMN to subcortical and frontal regions, which suggests that SZ appears less anatomically organized and structured with more extreme values, as also observed in the histogram Fig. \ref{fig:stat-z}-bottom.
\section{Conclusions and perspectives}
\label{sec:Conclusion}
In this paper, we presented the sparse representation of the subject-specific brain temporal functional network connectivity obtained from independent component analysis of the resting-state multi-subject fMRI dataset.
To this end, we suggested to jointly learn a dictionary for the sparse representation of the tFNC features and a linear classifier to determine whether the subjects should be classified as HC or Sz using sparse coefficients as features.
Compared with the FNC features, using sparse features, the classification rates improve.
More importantly, we identify new discriminative patterns formed from dictionary atoms that can be interpreted as tFNC features, \textit{i.e.}, revealing patterns of interaction between brain networks. 
%
This work also provides new perspectives for studying dynamics of fMRI to further investigate brain functionality.
Also, learning a non-linear classifier jointly with the dictionary can be used to further improve the classification rates \cite{mairal2008supervised}.

\newpage

\bibliographystyle{IEEEbib}
\small{\bibliography{biblio,strings,refs}
\balance

\end{document}